\documentclass[preprint2]{aastex}

\usepackage{hyperref}
\usepackage{amsmath} 
\usepackage{amsfonts} 
\usepackage{amssymb}
\usepackage{natbib}
\usepackage{subscript}
\usepackage[autostyle]{csquotes}

\shorttitle{Predicting CME arrivals with ElEvoHI}
\shortauthors{Rollett et al.}

\begin{document}


\title{ElEvoHI: a novel CME prediction tool for heliospheric imaging combining an elliptical front with drag-based model fitting}











\author{T.~Rollett$^1$, C.~M\"ostl$^{1,2}$, A.~Isavnin$^{3}$,  J.A.~Davies$^4$, M.~Kubicka$^1$, U.V.~Amerstorfer$^1$, R.A.~Harrison$^4$}

\altaffiltext{1}{Space Research Institute, Austrian Academy of Sciences, 8042 Graz, Austria}
\altaffiltext{2}{IGAM-Kanzelh\"ohe Observatory, Institute of Physics, University of Graz, 8010 Graz, Austria}
\altaffiltext{3}{Department of Physics, University of Helsinki, P.O. Box 64, 00014, Helsinki, Finland}
\altaffiltext{4}{RAL Space, Rutherford Appleton Laboratory, Harwell Campus, OX11 0QX, UK}





\begin{abstract}
In this study, we present a new method for forecasting arrival times and speeds of coronal mass ejections (CMEs) at any location in the inner heliosphere. This new approach enables the adoption of a highly flexible geometrical shape for the CME front with an adjustable CME angular width and an adjustable radius of curvature of its leading edge, i.e.\ the assumed geometry is elliptical. Using, as input, STEREO heliospheric imager (HI) observations, a new elliptic conversion (ElCon) method is introduced and combined with the use of drag-based model (DBM) fitting to quantify the deceleration or acceleration experienced by CMEs during propagation. The result is then used as input for the Ellipse Evolution Model (ElEvo). Together, ElCon, DBM fitting, and ElEvo form the novel ElEvoHI forecasting utility. To demonstrate the applicability of ElEvoHI, we forecast the arrival times and speeds of 21 CMEs remotely observed from STEREO/HI and compare them to in situ arrival times and speeds at 1 AU. Compared to the commonly used STEREO/HI fitting techniques (Fixed-$\phi$, Harmonic Mean, and Self-similar Expansion fitting), ElEvoHI improves the arrival time forecast by about $2$ hours to $\pm 6.5$ hours and the arrival speed forecast by $\approx 250$ km s$^{-1}$ to $\pm 53$ km s$^{-1}$, depending on the ellipse aspect ratio assumed. In particular, the remarkable improvement of the arrival speed prediction is potentially beneficial for predicting geomagnetic storm strength at Earth.
\end{abstract}


\keywords{Sun: coronal mass ejections (CMEs) --- Sun: heliosphere --- solar wind --- solar-terrestrial relations}

\section{Introduction}

Coronal mass ejections play a major role in space weather. These impulsive clouds of magnetized plasma have their origin in the solar corona and can reach Earth within a few days \citep[e.g.][]{schw05,she14}---fast CMEs can have transit times to 1~AU of less than a day \citep[e.g.][]{liu14}. 
Using numerical models, such as Enlil \citep[][]{ods03}, to propagate CMEs that have been characterized using coronagraph observations, leads to errors in CME arrival time predictions at Earth that lie in the range of $\pm 12$ to $\pm 18$ hrs \citep[e.g.][]{may15,vrs14}. Note that for a selected sample of CMEs \citet{mil13} obtained errors of $\pm 7.5$ hrs.
The reasons for these large forecasting errors are diverse. Firstly, the observations are limited. Currently, only the LASCO C2 and C3 coronagraphs \citep[][]{bru95} onboard the \textit{SOlar and Heliospheric Observat-
ory} (SOHO) and the COR1 and COR2 coronagraphs onboard the \textit{Ahead} spacecraft of the twin satellite mission \textit{Solar Terrestrial Relations Observatory} \citep[STEREO;][]{kai08} can be used to operationally forecast the arrival times of Earth-directed CMEs. 
Secondly, the structures, shapes, orientations, sizes, directions and speeds of CMEs are highly variable, i.e.\ it is quite difficult to describe all CMEs by a single propagation model.

Since the launch of STEREO, methods have been developed that exploit the observations made by the heliospheric imagers \citep[HI;][]{eyl09}, providing additional views of CMEs propagating all the way out to 1~AU and beyond. Many of those methods assume a specified geometry for the CME frontal shape (usually a circle subtending a fixed angular width at the Sun, which encompasses, in one limit, a point), a constant propagation speed and a fixed direction of motion. With these assumptions, it is possible to fit the time-elongation profile of the CME front to derive estimates of its launch time, radial speed and propagation direction from which its arrival time and speed at a specific target in interplanetary space, usually Earth, can be predicted \citep{she99,rou08,lug09b,lug10,moe11,dav12,moedav13}. \citet{moe14} applied such forecasting methods to 24 CMEs observed by STEREO/HI and found a mean difference between the forecasted and detected arrival times at 1~AU of $8.1 \pm 6.3$ hrs and a mean difference between forecasted and detected arrival speeds of $284 \pm 288$ km s$^{-1}$. The main drawback of these fitting methods is the constant speed assumption that systematically overestimates the arrival speed, especially of CMEs that are actually decelerating \citep[][]{lugkin12}. The propagation speed of CMEs tends to approach that of the ambient solar wind, i.e.\ fast events tend to decelerate and slow ones tend to accelerate \citep[e.g.][]{vrs04}. This is ongoing through the STEREO/HI1 field of view \citep[][]{tem11,rol14}, which extends from $4^\circ$ to $24^\circ$ elongation, the latter of which corresponds to radial distances of $\approx 100$ R$_\odot$.
In order to be able to account for such an evolution in CME speed, the drag-based model was developed \citep[DBM;][]{vrs13} \textbf{and} has already been used in a multitude of studies \citep[e.g.][]{tem11,tem12,mis13,rol14,temnit15,zic15}.
As an extension to the DBM, \cite{moe15} developed the Ellipse Evolution Model (ElEvo). It assumes an elliptically shaped CME front and also can be used to predict arrival times and speeds at specified locations in space. The disadvantage of ElEvo and DBM is that they rely on coronagraph data, which allows CMEs to be observed out to a maximum heliocentric distance of only $32$ R$_{\odot}$. Heliospheric imagery provides the possibility of tracking a CME out to a much larger distance leading to the chance of achieving better reliability of its derived kinematics.

In this study, we present a new method of exploiting single spacecraft HI observations, either from STEREO or from any other future mission carrying such instrumentation, such as the Wide-Field Imager for \textit{Solar Probe Plus} \citep[WISPR;][]{vou15} or the heliospheric imager \citep[SoloHI;][]{how13a} aboard \textit{Solar Orbiter}: the Ellipse Conversion method (ElCon). ElCon converts the observed elongation angle (the angle between the Sun-observer line and the line of sight) into a radial distance from Sun center, assuming an elliptical CME front propagating along a fixed direction. The CME width, propagation direction and the ellipse aspect ratio are free parameters. 
The combination of the ElCon method, the DBM fitting, and the ElEvo model allows use of HI observations to forecast CME arrival time without compromising arrival speed; this combination forms the new forecasting utility ElEvoHI.

\section{Data}
\label{sec:data}

In order to introduce and test ElEvoHI, we analyze a sample of 21 CMEs observed by HI on the STEREO A spacecraft between the years 2008 and 2012. For each event, the in situ arrival time and speed of the CME at 1~AU is available, either from its passage over \textit{Wind} or over STEREO~B. The sample covers slow events during the solar minimum period as well as fast CMEs during solar maximum. The list was compiled by \citet{moe14} to test existing HI fitting methods used for forecasting CME arrival times and speeds.
While the event list of \citet{moe14} contains 24 CMEs, this study excludes three of them, for different reasons. Event number 3 \citep[from][]{moe14} was the trailing edge of a CME. We exclude this event since our study focuses on predicting the arrival of the CME leading edge. Event number 4 was imaged from STEREO B, while in this study we consider only CMEs that were imaged from STEREO A. Finally, we exclude event number 11, which is the same CME as event number 12, but in situ detected by STEREO B.
 
\section{ElEvoHI: The Ellipse Evolution Model based on HI observations}

The Ellipse Evolution model (ElEvo) was developed by \citet{moe15} and assumes an elliptical CME leading edge with a predefined half-width and aspect ratio. It makes use of the DBM \citep[][]{vrs13} and is able to provide forecasts of arrival time and speed at any target in the inner heliosphere.

Altogether, ElEvo needs 8 input parameters: the ellipse angular half-width, $\lambda$, and inverse aspect ratio, $f$, the propagation direction, $\phi$, the start time, $t_{\rm init}$, and initial speed, $v_{\rm init}$, the latter two at the radial distance, $r_{\rm init}$, the mean background solar wind speed, $w$, and the drag parameter, $\gamma$.
Using the newly developed ellipse conversion method (ElCon, see Appendix) in combination with Fixed-$\phi$ fitting \citep[FPF;][]{she99,rou08}, as discussed below, it is possible to estimate all of these parameters (barring $\lambda$ and $f$) based on heliospheric imager observations only, consistent with all the requirements of ElEvo; we call this HI-based alternative to ElEvo, ElEvoHI.

\begin{figure}[!htbp]
\plotone{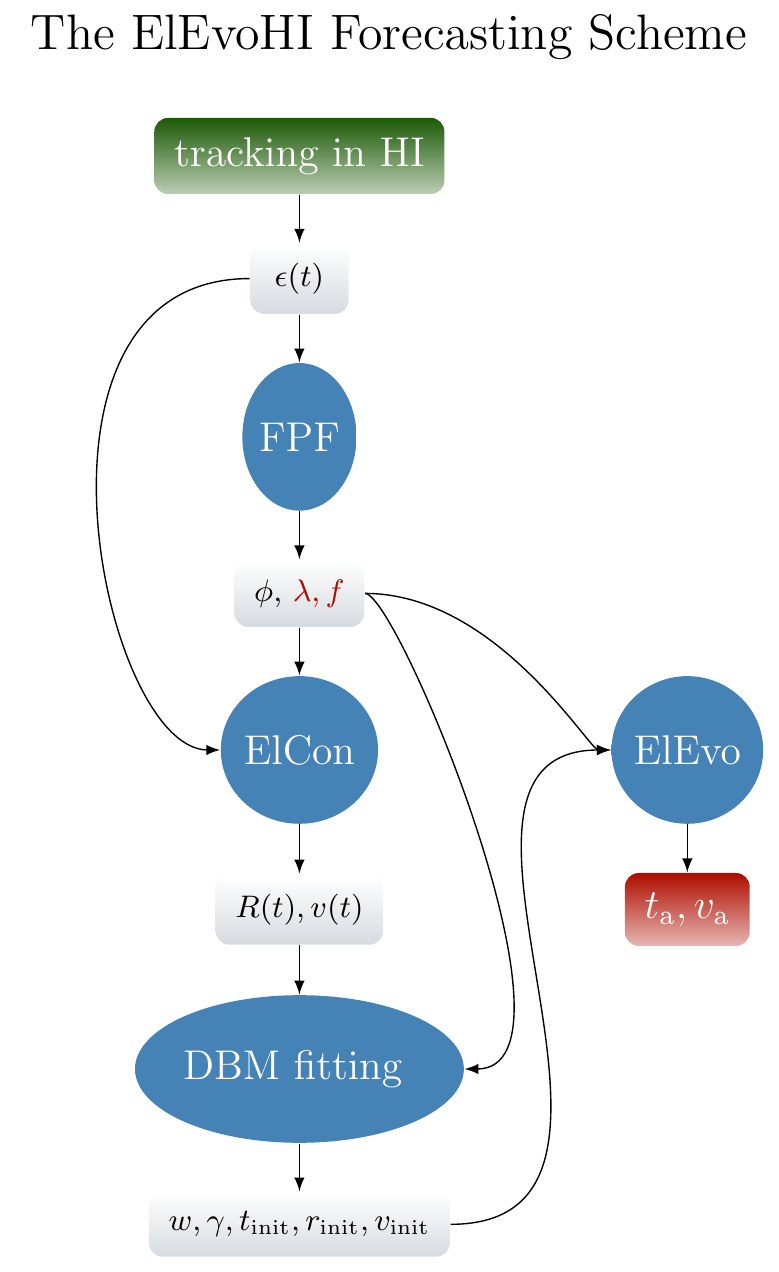}
\caption{Flowchart to visualize all of the constituent methods that combine to produce ElEvoHI. The blue ellipses symbolize the components used, and the grey rectangles show their output parameters, many of which are input into ElEvo, providing the forecast of the arrival time, $t_{\rm a}$, and speed, $v_{\rm a}$. \label{fig:fchart}}
\end{figure}

The required parameters can be provided by a combination of ElCon and DBM fitting. It is possible to derive $\phi$ from the FPF technique, which is an easy and fast approach applied to HI elongation profiles, i.e.\ no other data is needed. Using only STEREO/HI data, we need to assume $\lambda$ and $f$. Figure \ref{fig:fchart} shows the forecasting scheme of ElEvoHI. The blue ellipses show the different components of the method, the grey parts show the parameters required by, and obtained from, each of these components. Starting at the top of the flowchart, we acquire the time-elongation profile, $\epsilon (t)$, from HI observations. FPF analysis of the time-elongation profile provides an estimate of $\phi$ as input for ElCon, which then converts the elongation profile to radial distance assuming an elliptical geometry (see Appendix). As noted previously, $\lambda$ and $f$, also required as input to ElCon, must be assumed. The distance profile produced by ElCon is then fitted using the DBM---building the derivative of both yields the speed profile. The required parameters are then input into ElEvo, which forecasts the arrival time, $t_{\rm a}$, and the arrival speed, $v_{\rm a}$, of the CME. As noted above, we term this entire procedure ElEvoHI.

\subsection{The Ellipse Conversion Method (ElCon)}

The first step in forecasting CME arrival time and speed using ElEvoHI is to track the time-elongation profile of the CME in the ecliptic plane. For this purpose the SATPLOT tool\footnote{\url{http://hesperia.gsfc.nasa.gov/ssw/stereo/secchi/idl/jpl/satplot/SATPLOT_User_Guide.pdf}} is very convenient. With this tool we can extract the CME track from a time-elongation map \citep[commonly called a J-map;][]{dav09b} and, additionally, a FPF analysis can be performed. Time-elongation profiles for all CMEs, presented by \citet{moe14} and used in this article, are extracted using the SATPLOT tool.

Similar to the three conversion methodologies, well used for interpreting STEREO/HI data, based on Fixed-$\phi$ \citep[][]{she99,kahweb07}, Harmonic Mean \citep[HM;][]{howtap09,lug09b} and Self-similar Expansion \citep[SSE;][]{dav12} geometries, ElCon can be used to convert elongation to radial distance. The mathematical derivation of the ElCon method is given in the Appendix. By applying Equation \ref{eq:rell} from the Appendix, the observed elongation angle along with a CME is detected, $\epsilon$, can be converted into the radial distance, $R_{\rm ell}$, of the CME apex from Sun center. It is assumed that the line-of-sight from the observer forms the tangent to the leading edge of the CME, similar to the HM and SSE conversion methods. 

\begin{figure}[!htbp]
\epsscale{1}
\plotone{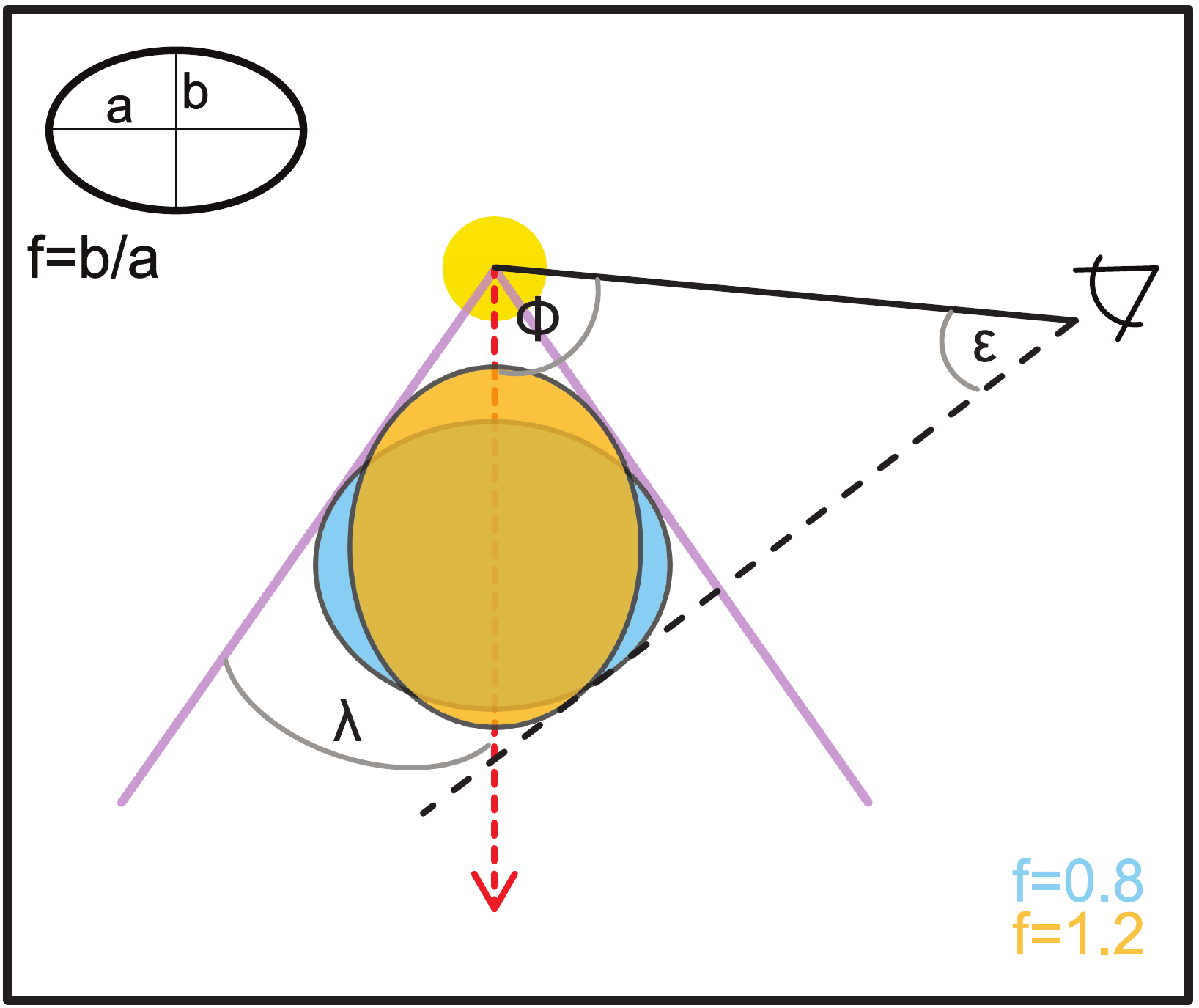}
\caption{Examples of two different elliptic CME fronts. Both ellipses conform to the same parameter set ($\lambda$, $\phi$, $\epsilon$) but have different inverse aspect ratios, $f$.\label{fig:elcon}}
\end{figure}

Figure \ref{fig:elcon} illustrates two elliptically shaped CME fronts with different inverse aspect ratios, namely $f=0.8$ (blue ellipse) and $f=1.2$ (orange ellipse). In this depiction, both CME fronts are observed at the same elongation angle, $\epsilon$, in the same propagation direction, $\phi$. Moreover, both have the same angular half-width, $\lambda$. From this figure it is clear that the time of impact of the CME at the in-situ observatory would not only depend upon its radial speed, but would also clearly depend critically upon the CME’s inverse aspect ratio (as well, of course, as the angular offset between the in situ observatory and the CME apex). Moreover, it is also clear, that for a CME where $b>a$ (where $a$ is the semi-major axis and $b$ is the semi-minor axis) it is particularly important to have an accurate propagation direction in order to achieve an accurate arrival time.

\subsubsection{Input for ElCon}
\label{sec:input}

For converting the elongation of the CME in the STEREO/HI observations to radial distance, we need to input the assumed half-width, $\lambda$, and inverse aspect ratio, $f=b/a$, and the propagation direction, $\phi$. Here, we obtain the latter parameter by fitting the time-elongation profile from STEREO/HI using the FPF method.

\subsubsubsection{Fixed-$\phi$ Fitting Method}

The Fixed-$\phi$ fitting method \citep[FPF;][]{she99,rou08} is a commonly used, fast and easy method to predict arrival times of CMEs from heliospheric imagery. Although it has some major disadvantages, in particular the point-like shape and constant radial speed assumed for the CME, it does not, in fact, show a larger error than more sophisticated methods assuming an extended shape for the CME front \citep[see][]{moe14}. 
To obtain the propagation direction from FPF, \citet{moe14}, whose results we use here, fitted the time-elongation track up to $40^\circ$ elongation. Note that FPF uses the same data as ElCon, i.e.\ no additional data is needed. This is a big advantage of FPF over other potential methods, such as the Graduated Cylindrical Shell model \citep[GCS;][]{the09} for determining the input value of $\phi$ for ElCon. The disadvantage when using FPF is, that we have to assume the input parameters $\lambda$ and $f$ for ElCon. Other methods, such as GCS, could potentially provide estimates of $\lambda$ and $f$.

The equation to fit the time-elongation profile is given by

\begin{equation}
\epsilon(t) = \arctan \left(\frac{R(t) \sin(\phi)}{d_{\rm o}-R(t) \cos(\phi)}\right),
\label{eq:fpf}
\end{equation}

where $d_{\rm o}$ is the distance between Sun center and the observer.

\begin{figure}[!htbp]
\epsscale{0.9}
\plotone{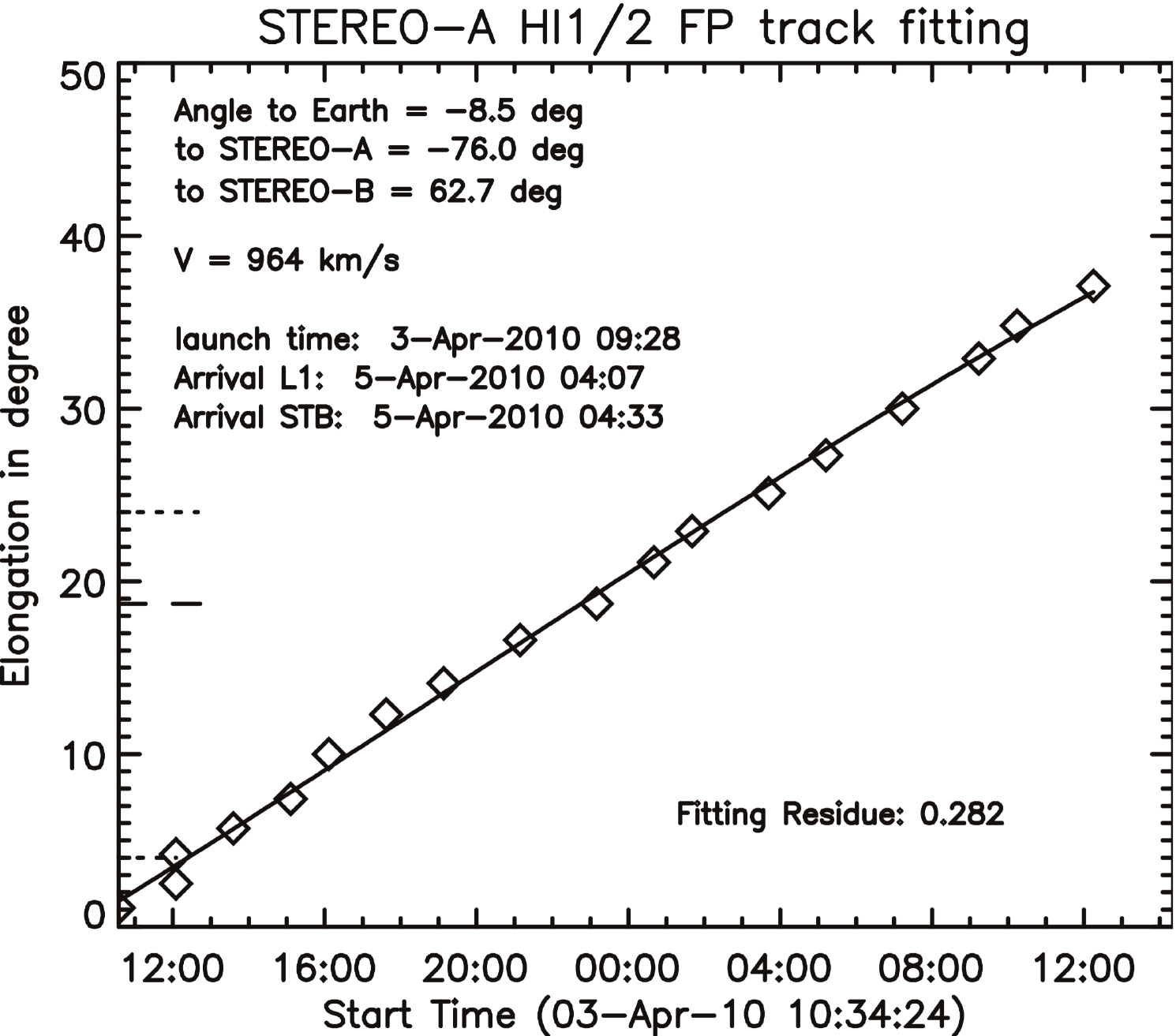}
\caption{Example of a time-elongation fit using FPF, which is implemented within the SATPLOT tool. The diamonds show the tracked points and the solid line shows the best FPF fit to those points. The propagation direction, the speed as well as the arrival times at 1 AU are given. The fitting residue is quoted in degrees of elongation. \label{fig:fpf}}
\end{figure}

Figure \ref{fig:fpf} shows an example of STEREO/HI time-elongation profile (diamonds) manually tracked using SATPLOT \citep[from][their Event number 7]{moe14}. The FPF, HMF and SSEF methods can also be applied within SATPLOT, the solid line shows the best FPF fit to the tracked points. The event displayed is Event number 5 in Table \ref{tab:results}.

Having obtained $\phi$ from FPF as an input parameter for the ElCon conversion method \citep[as previously noted, we use values from][]{moe14}, we can apply ElCon to the HI time-elongation profile. As noted above, we need to assume $\lambda$ and $f$.
ElCon yields the radial distance profile of the ellipse apex from Sun center, $R_{\rm ell}$, and the corresponding speed profile as input for DBM fitting.

\subsection{Connecting with the Drag-Based Model}

After converting the STEREO/HI elongations to radial distances by applying ElCon, we fit the ElCon time-distance profile using the DBM developed by \citet{vrs13}. The DBM considers the influence of the drag force acting on solar wind transients during their propagation through interplanetary space. It is based on the assumption that, beyond a distance of about 15~R$_\odot$ from the Sun, the driving Lorentz force can be neglected and the drag force can be considered as the predominant force affecting the propagation of a CME \citep[][]{vrs06}. Under these circumstances, the equation of motion of a CME can be expressed as

\begin{equation}
R_{\rm DBM}(t) = \pm \frac{1}{\gamma} \ln [1 \pm \gamma (v_{\rm init}-w) t] + w t + r_{\rm init},
\label{eq:dbm}
\end{equation}
where $R_{\rm DBM}(t)$ is the radial distance from Sun-center, $\gamma$ is the drag parameter (usually ranging from $0.2\times10^{-7}$ km$^{-1}$ to $2\times10^{-7}$ km$^{-1}$), $v_{\rm init}$ and $r_{\rm init}$ are the initial speed and distance, respectively, and $w$ is the background solar wind speed. The sign $\pm$ is positive when $v_{\rm init} > w$ and negative when $v_{\rm init} < w$.

\begin{figure}[!htbp]
\epsscale{1.1}
\plotone{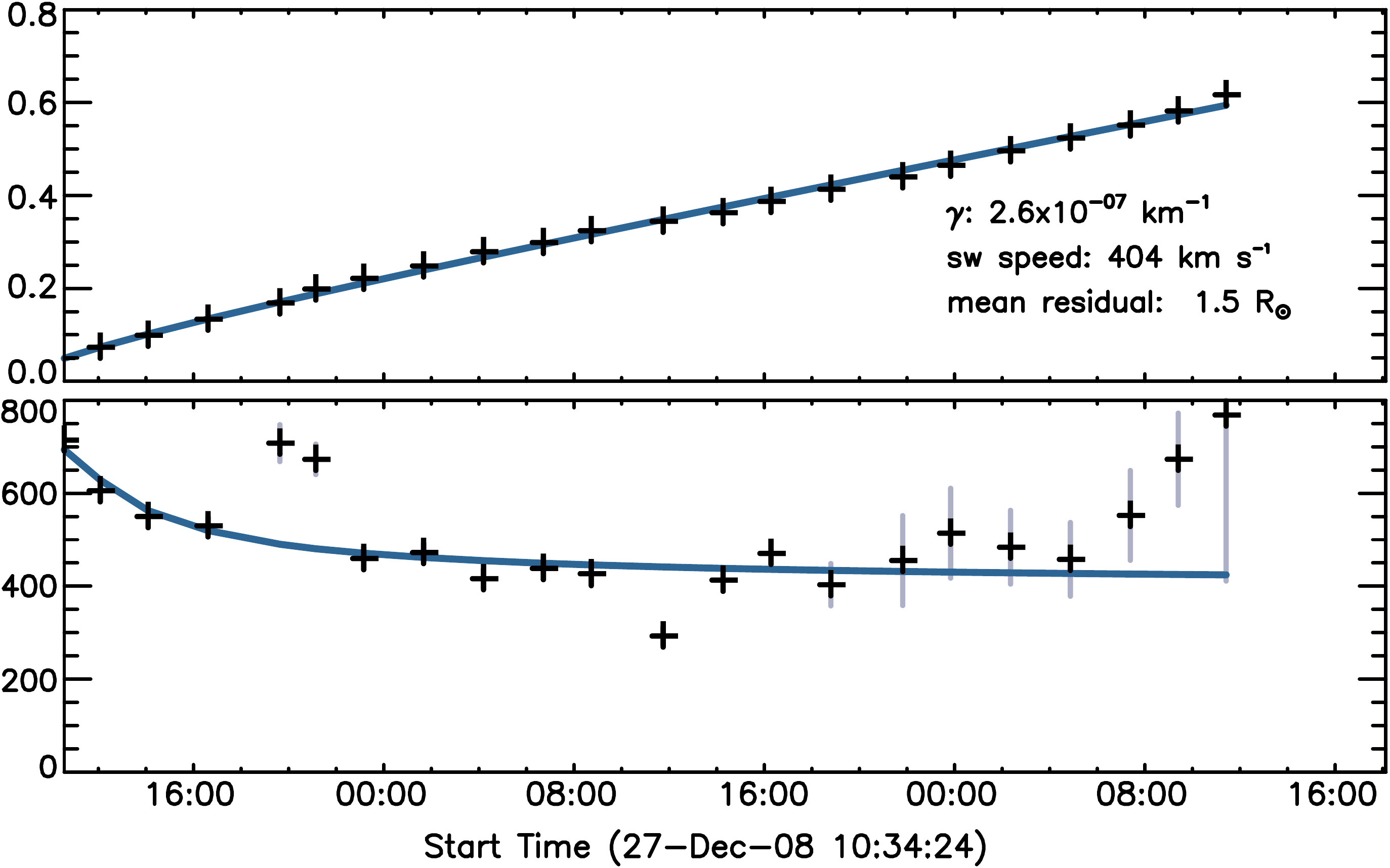}
\caption{HI measurements of the time-elongation profile of a CME converted to a time-heliocentric distance profile using ElCon (crosses). The vertical lines show the standard deviation resulting from the HI measurements. The blue line shows the DBM fit, which (in our implementation) provides estimates of the drag-parameter and the background solar wind speed. \label{fig:dbmf}}
\end{figure}

\citet{zic15} implemented a DBM fitting technique, in which a solar wind model proposed by \citet{leb98} was used to provide parameter $w$ as an input to the model. In contrast to that work, our version of the DBM fitting produces the best-fit value of $w$ as a quasi-output, as its value is constrained by input in situ measurements (as discussed below). The mean, $w_{\rm mean}$, minimum, $w_{\rm min}$, and maximum, $w_{\rm max}$, values of the in situ solar wind speed at the detecting spacecraft (either STEREO or \textit{Wind}), over the same time range as the remote observations, are used to define a range of possible values for $w$. Using these values, five different DBM fits are performed, for $w \in \{w_{\rm min},$ $w_{\rm min}+(w_{\rm mean}-w_{\rm min})/2$, $w_{\rm mean}$, $w_{\rm mean}+(w_{\rm max}-w_{\rm mean})/2$, $w_{\rm max}\}$. The value of $w$ that yields the fit with the smallest residuals, defines the value of $w$ used.

The drag parameter, $\gamma$, is also output from the DBM. This parameter is a combination of various properties of the CME and can be expressed as $\gamma = c_{\rm d} A \rho_w/m$, where $c_{\rm d}$ is the dimensionless drag coefficient, $A$ is the cross section area of the CME, $\rho_w$ is the solar wind density, and $m$ is the CME mass \citep[][]{car04}. Note that $\gamma$ is, however, fitted as a single parameter.

Figure \ref{fig:dbmf} shows an example of a time-distance profile in units of AU (upper panel), resulting from the application of ElCon to a HI1/HI2 time-elongation profile of CME number 5 from \citet{moe14} (black crosses); the lower panel shows the corresponding speed profile of the CME apex, which is obtained by differentiating the ElCon time-distance profile. The light blue vertical lines in the lower panel mark the standard deviation in the velocity resulting from a measurement error of $0.1^\circ$ (for HI1) and $0.4^\circ$ (for HI2) in elongation \citep{rol13}. These elongation errors, similar to the measurements themselves, are converted to a distance error (and subsequently to a speed error) using ElCon. The errors in the time-distance profile are so small as to be not visible. The blue curve in the upper panel of Figure \ref{fig:dbmf} represents the DBM time-distance fit. The DBM speed profile, the blue line in the lower panel, is obtained by differentiating the DBM time-distance fit.

One difficulty, when applying the DBM fit to the ElCon output, is defining the starting time, $t_{\rm init}$, and the corresponding starting distance, $r_{\rm init}$, of the fit. The DBM only considers forces akin to \enquote*{aerodynamic} drag, so it is only really valid over the altitude regime covered by the HI observations. Note that COR2 data is included in SATPLOT as well as HI data. The best value of $r_{\rm init}$ for the DBM fit is chosen as that point that yields the best overall fit, i.e.\ that which gives the smallest residuals. This varies for each event, and for every combination of assumed angular width and assumed aspect ratio. The average starting point for the DBM fit in our sample of CMEs lies at $21 \pm 10$ R$_\odot$. Depending on $t_{\rm init}$ and $r_{\rm init}$, the value of $v_{\rm init}$ is taken from the ElCon speed-profile corresponding to $t_{\rm init}$. The mean value of the fitting residuals over all 21 CME fitted lies between 1.5 and 1.8 R$_\odot$.

By applying the DBM fitting procedure, we acquire all input parameters to conduct the last step of ElEvoHI, namely to run ElEvo, which provides the forecast of arrival time and speed.

\subsection{Running ElEvo}

\begin{figure}[!htbp]
\epsscale{1}
\plotone{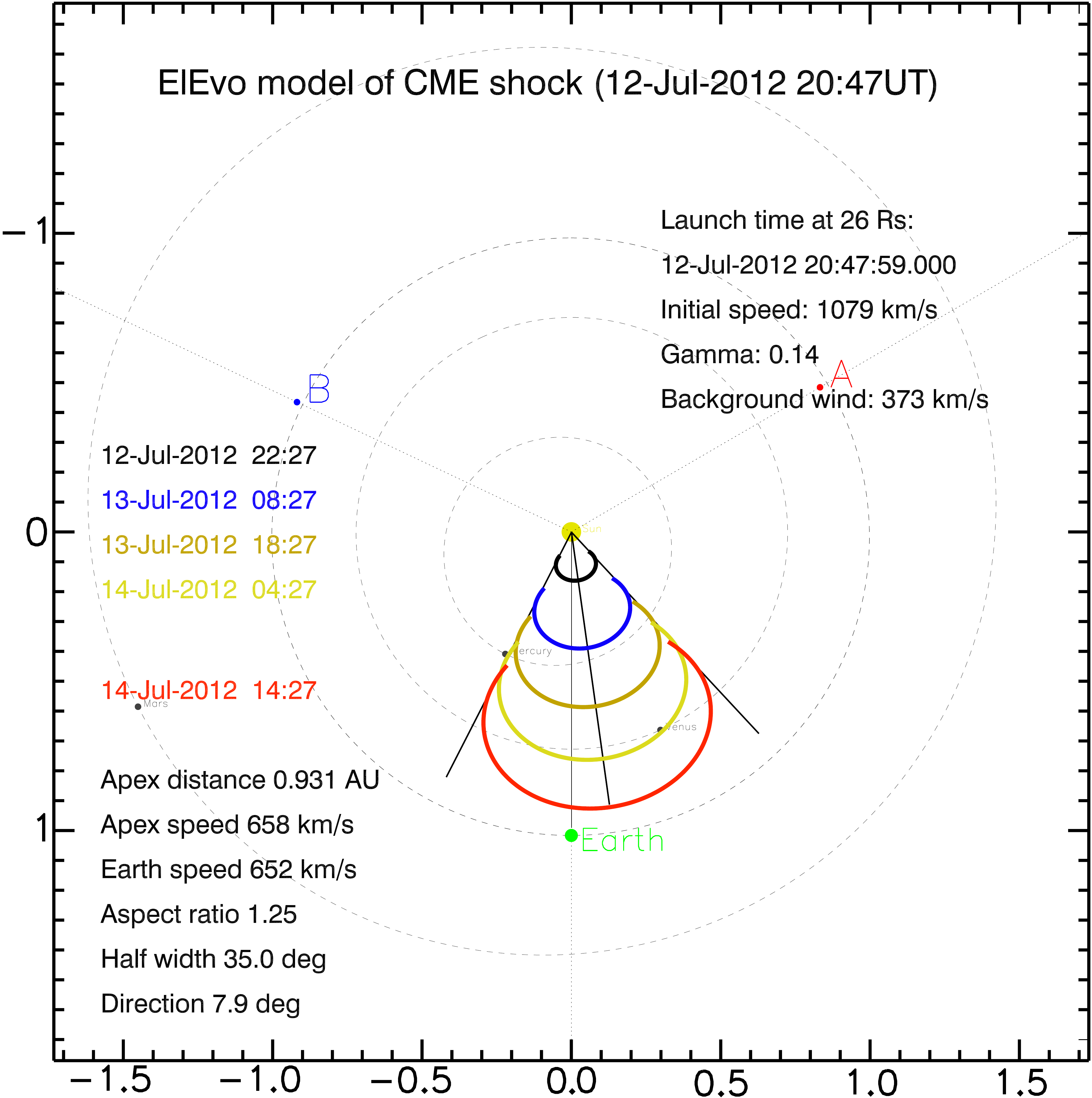}
\caption{Visual output of ElEvo with different time steps labeled in different colors, axes are in HEE coordinates. In this example $f=0.8$ and $\lambda=35^\circ$.\label{fig:elevo}}
\end{figure}

The ElEvo model can be used to predict CME arrival times and speeds at any specified point in interplanetary space (usually the location of a spacecraft making in situ solar wind measurements). ElEvo assumes the same elliptical geometry as ElCon and includes the DBM to simulate the propagation of the CME beyond the extent of the observations. In the past, ElEvo has been run based on coronagraph data \citep[][]{moe15}. 
To run ElEvo, one needs to know the following CME parameters: $\phi$, $\lambda$, $f$, $t_{\rm init}$, $r_{\rm init}$ and $v_{\rm init}$ as well as $\gamma$ and $w$. The latter five parameters are, as explained above, gained through the combination of ElCon and DBM fitting. The propagation direction, $\phi$, results from FPF. Use of FPF means that $\lambda$ and $f$ have to be assumed. Figure \ref{fig:elevo} shows an example of an ElEvo run. Different times during the propagation of the CME are plotted in different colors. All parameters input to ElEvo for this CME (Event number 20 in Table \ref{tab:results}) are written in the upper right and lower left corners of the figure. For the time of the red colored front, the speed of the CME apex and the CME speed in the direction of the in situ observatory are also marked on the panel.

\section{Test of ElEvoHI Forecast on 20 CMEs}
\label{sec:20cmes}
\begin{figure}[!htbp]
\epsscale{1.0}
\plotone{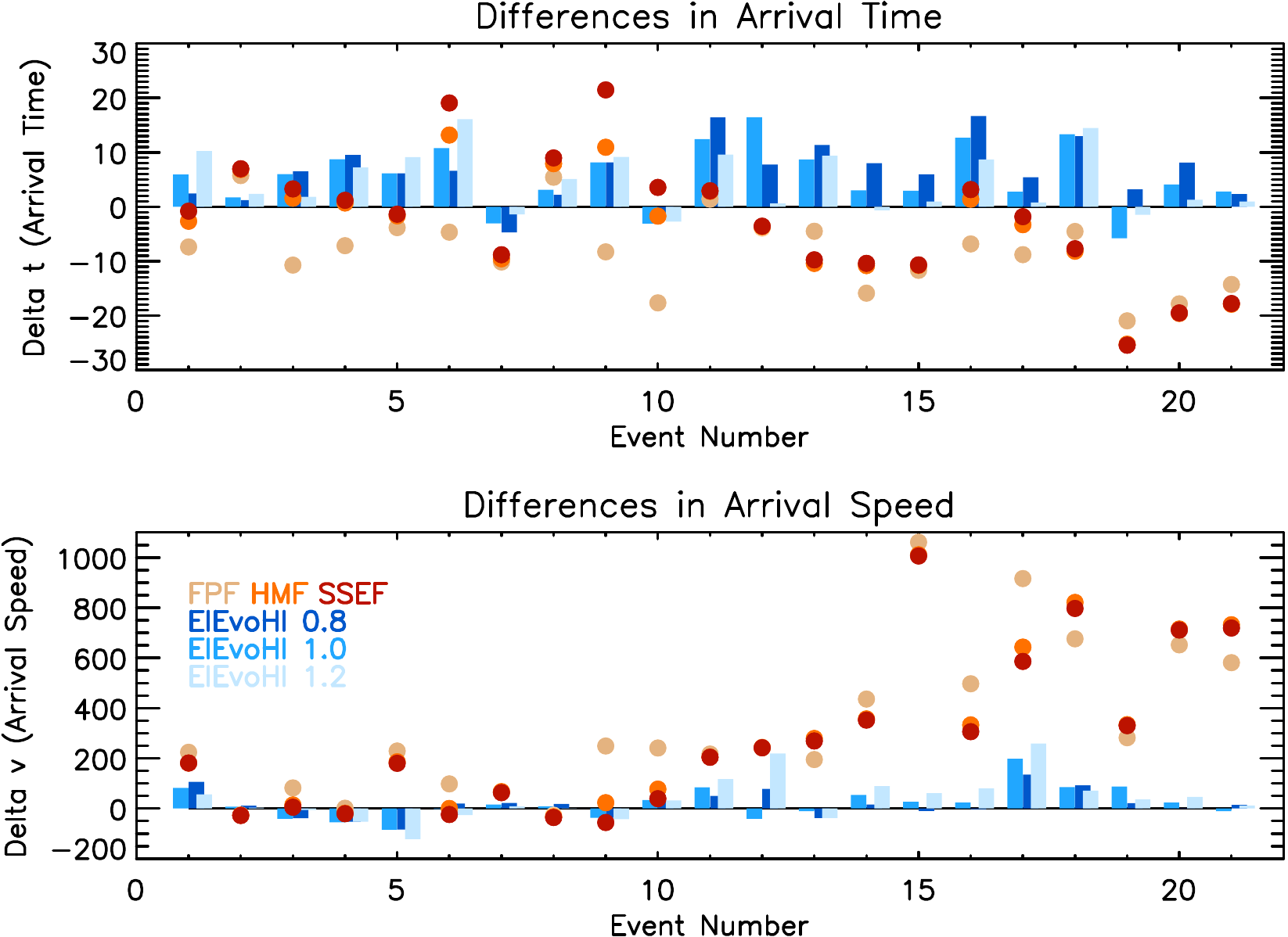}
\caption{Comparison of ElEvoHI, FPF, HMF and SSEF forecasts. The upper panel shows the differences between the predicted and in situ arrival times ($\langle \Delta t \rangle = \langle t_{\rm ElEvoHI} -t_{\rm obs} \rangle$), the lower panel illustrates the equivalent plot for the arrival speed ($\langle \Delta v \rangle = \langle v_{\rm ElEvoHI}- v_{\rm obs} \rangle$). The salmon (orange, red) colored circles correspond to the FPF (HMF, SSEF) forecast, the bluish bars correspond to ElEvoHI with the three different aspect ratios used. \label{fig:forecomp}}
\end{figure}

In our application of ElEvoHI to 21 of the 24 CMEs previously analyzed by \citet{moe14}, we have used three different inverse aspect ratios ($f \in \{0.8, 1, 1.2\}$) to test and assess the performance of ElEvoHI. Note that $f=1$ corresponds to the SSE (circular) geometry with the same angular width. As noted above, as input propagation direction for each CME, we use the corresponding FPF values from \citet{moe14}, who tracked the CMEs up to about $40^\circ$ elongation. Note that the Fixed-$\phi$ geometry corresponds to the SSE and ElCon geometries with $\lambda=0$. For all CMEs, we use the same value of the half-width ($\lambda=35^\circ$). In reality, of course, every CME is different and one would not expect them to have the same half-width, or indeed aspect ratio. As we are only introducing and testing ElEvoHI here, we have decided to keep our analysis as simple as possible and have hence fixed the half-width to $35^\circ$. Following studies will no doubt use different values for the half-width. As mentioned above, using the Graduated Cylindrical Shell model \citep[GCS;][]{the09}, based on coronagraph or even HI data \citep[][]{col13}, it is possible to derive $f$ and $\lambda$ individually. In Table \ref{tab:taball} in the Appendix the resulting fitting parameters for each event and the three half-widths tested are given.

\begin{deluxetable}{llrcrrrrrr}
\tablecolumns{10}
\tablewidth{0pc}
\tabletypesize{\scriptsize}
\tablecaption{Differences between ElEvoHI predictions and in situ arrival time and speed \label{tab:results}}
\tablehead{
\multicolumn{3}{c}{Event} & \colhead{} & \multicolumn{6}{c}{ElEvoHI}\\
\cline{1-3} \cline{5-10} \\
\colhead{n\textordmasculine} & \colhead{$t^1_{\rm obs}$} & \colhead{$v^1_{\rm obs}$} & \colhead{} & \colhead{$\Delta t^{2}_{0.8}$} & \colhead{$\Delta v^{3}_{0.8}$}& \colhead{$\Delta t^{2}_{1}$} & \colhead{$\Delta v^{3}_{1}$}& \colhead{$\Delta t^{2}_{1.2}$} & \colhead{$\Delta v^{3}_{1.2}$}}
\startdata
1 & 2008 Apr 29 13:21 & 430 & & 2.42 & 106 & 5.92&82 & 10.25& 56\\
2 & 2008 Jun 6 15:35 & 403   & & 1.22 &  11 &1.72 &7 & 2.38& 3\\
3 & 2008 Dec 31 01:45 & 447 & & 6.48 & $-39$ &5.98 &$-41$ & 1.82 & $-6$ \\
4 & 2009 Feb 18 10:00 & 350& & 9.53 & $-53$ &8.73 & $-55$& 7.23 & $-53$\\
5 & 2010 Apr 5 07:58 & 735 & & 6.13 & $-84$ &6.13 & $-85$& 9.13 & $-122$\\
6 & 2010 Apr 11 12:14 & 431 & & 6.62 & 20 &10.78 &$-2$ & 16.12 & $-26$\\
7 & 2010 May 28 01:52 & 370 & & $-4.73$ & 22 &$-3.06$ &16 & $-1.40$ & 9\\
8 & 2010 Jun 20 23:02 & 400 & & 2.20 & 18 &3.12 & 9& 5.12 &$-1$\\
9 & 2010 Aug 3 17:05 & 581 & & 8.15 & $-37$ &8.15 & $-37$& 9.15 &$-43$\\
10 & 2011 Feb 18 00:48 & 497 & & $-2.11$ & 28 &$-3.10$ & 34& $-2.77$ &32\\
11 & 2011 Aug 4 21:18 & 413  & & 16.43 & 50 & 12.43& 84& 9.60 & 117\\
12 & 2011 Sep 9 11:46 & 489  & & 7.77 & 78 &16.47 & $-41$& 0.60 & 219\\
13 & 2011 Oct 24 17:38 & 503 & & 11.37 & $-38$ &8.70 & $-11$& 9.38 & $-38$\\
14 & 2012 Jan 22 05:28 & 415& & 8.00 & 16 &3.00 & 54& $-0.67$ & 89\\
15 & 2012 Jan 24 14:36 & 638  & & 5.95 & $-10$ &2.95 & 27& 0.95 & 61\\
16 & 2012 Mar 7 03:28 & 501 & & 16.68 & $-3$ &12.68 & 24& 8.68 & 80\\
17 & 2012 Mar 8 10:24 & 679 & & 5.43 & 135 &2.77 & 198& 0.77 & 259\\
18 & 2012 Mar 12 08:28 & 489 & & 12.98 & 92 &13.32 & 85& 14.48 & 71\\
19 & 2012 Apr 23 02:14 & 383 & & 3.18 & 21 &$-5.80$ & 87& $-1.47$ & 36\\
20 & 2012 Jun 16 19:34 & 494 & & 8.10& $-1$  & 4.10& 24& 1.27 & 46\\
21 & 2012 Jul 14 17:38 & 617 & & 2.32 & 14 &2.78 & $-11$& 0.95 & 12\\
\enddata
\tablenotetext{1}{In situ arrival times and speeds taken from \citet{moe14}.}
\tablenotetext{2}{$\langle \Delta t \rangle = \langle t_{\rm ElEvoHI} -t_{\rm obs} \rangle$ in hrs}
\tablenotetext{3}{$\langle \Delta v \rangle = \langle v_{\rm ElEvoHI} -v_{\rm obs} \rangle$ in km s$^{-1}$}
\end{deluxetable}

In order to assess the ability of ElEvoHI to improve the accuracy of predicting CME arrival times and speeds, we compare the outcome to the commonly used HI fitting methods, FPF, Harmonic Mean fitting \citep[HMF;][]{lug11} and Self-similar Expansion fitting \citep[SSEF;][]{moedav13}. HMF assumes a circular CME front with a half-width of $90^\circ$, which means that the circle is always attached to the Sun. SSEF assumes a circular CME frontal shape as well but the half-width is variable. For the events in the list, \citet{moe14} have set $\lambda=45^\circ$.

Figure \ref{fig:forecomp} shows the resulting forecasts of the arrival times and speeds of all 21 CMEs considered in this study, for the three different values for $f$ used, along with the FPF, HMF and SSEF predictions. The upper panel shows the differences in arrival time ($\langle \Delta t \rangle = \langle t_{\rm ElEvoHI} -t_{\rm obs} \rangle$), where values $< 0$ indicate that a CME was predicted to arrive earlier than it actually arrived in situ (at 1 AU) and values $> 0$ indicate that a CME was predicted to arrive after it actually arrived. While the FPF, HMF and SSEF techniques (plotted using yellow, orange and red circles, respectively) tend to predict CME arrival too early ($\langle\Delta t\rangle_{\rm FPF} = -7.9 \pm 7.1$ hrs, $\langle\Delta t\rangle_{\rm HMF} = -3.8 \pm 10$ hrs, $\langle\Delta t\rangle_{\rm SSEF} = -2.3 \pm 11.6$ hrs), ElEvoHI (light, medium and dark blue bars) nearly always predicts CME arrival too late. The best ElEvoHI arrival time forecast is found using $f=1.2$, with $\langle \Delta t \rangle = 5.0 \pm 5.6$ hrs (light blue bars). Using $f=1$ (equivalent to the SSE geometry) results in $\langle \Delta t \rangle = 5.7 \pm 5.9$ hrs (medium blue bars), using $f=0.8$ leads to $\langle \Delta t \rangle=6.3 \pm 5.5$ hrs (dark blue bars).
The lower panel of Figure \ref{fig:forecomp} shows the equivalent plot for the arrival speed ($\langle \Delta v \rangle = \langle v_{\rm ElEvoHI}- v_{\rm obs} \rangle$). For our sample of CMEs, at least, ElEvoHI provides a substantial improvement in forecasting CME speed when compared to the FPF, HMF and SSEF techniques, for which $\langle \Delta v \rangle_{\rm FPF} = 328 \pm 302$ km s$^{-1}$, $\langle \Delta v \rangle_{\rm HMF} = 292 \pm 314$ km s$^{-1}$ and $\langle \Delta v \rangle_{\rm SSEF} = 277 \pm 314$ km s$^{-1}$. The mean difference between the forecasted and observed in situ arrival speeds is least for $f=0.8$, with $\langle \Delta v \rangle = 17 \pm 54$ km s$^{-1}$. Using $f=1$, we find $\langle \Delta v \rangle = 21 \pm 63$ km s$^{-1}$ and using $f=1.2$ gives $\langle \Delta v \rangle = 38 \pm 87$ km~s$^{-1}$. Table \ref{tab:results} lists all analyzed events and quotes $\Delta t$ and $\Delta v$ for each of the three inverse aspect ratios used in this study.

\begin{figure}[!htbp]
\plotone{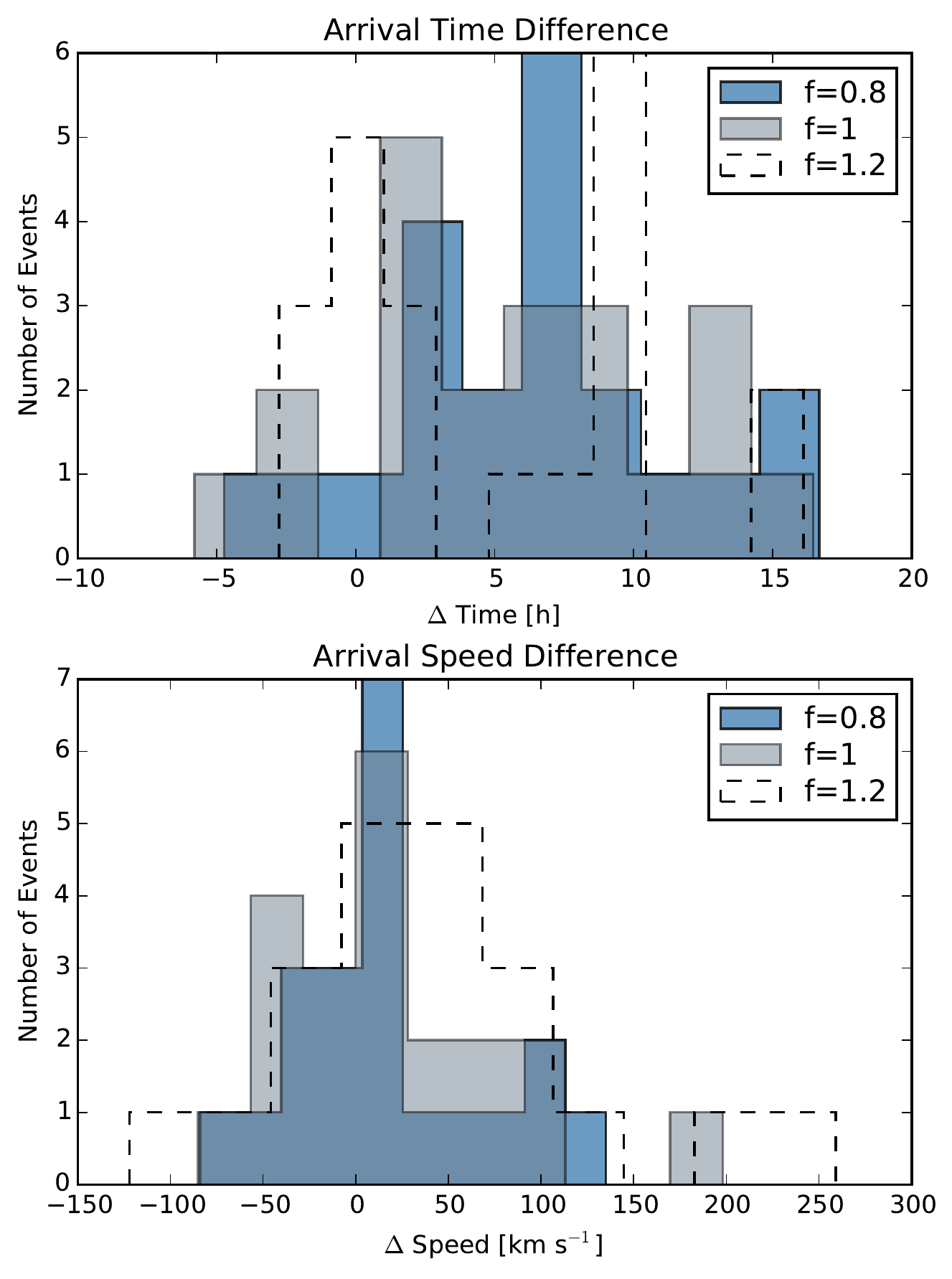}
\caption{Frequency distributions of the differences between predicted and observed arrival times (top) and speeds (bottom). The different colors correspond to the three different inverse aspect ratios used for ElEvoHI. \label{fig:histogram}}
\end{figure}

Figure \ref{fig:histogram} shows frequency distributions of $\Delta t$ (upper panel) and $\Delta v$ (lower panel) for ElEvoHI. The blue, grey, and white areas (the latter bounded by a dashed line) represent $f=0.8$, $f=1$, and $f=1.2$, respectively. Regardless of which aspect ratio is used, all arrival time forecasts lie within the range $-5.8$~hrs~$ < \Delta t < 16.7$ hrs, compared to FPF ($-21$~hrs~$ < \Delta t < 5.7$~hrs), HMF ($-25$~hrs~$ < \Delta t < 13$~hrs) and SSEF ($-25$~hrs~$ < \Delta t < 21$~hrs). The minimum and maximum values of $\Delta v$ for the ElEvoHI speed prediction are $-122$ and $259$ km s$^{-1}$, respectively. Because of the simplistic assumption of constant speed invoked by FPF, HMF and SSEF, the values of $\Delta v$ from these methods are much greater, ranging from $-27$ to $1061$ km s$^{-1}$ for FPF, from $-33$ to $1011$ km s$^{-1}$ for HMF and from $-56$ to $1006$ km s$^{-1}$ for SSEF. The average root mean square values of $\Delta t$ and $\Delta v$ from ElEvoHI, over all aspect ratios used, are $5.4$ hrs and $66$ km s$^{-1}$, respectively. The corresponding values for FPF, HMF and SSEF are $7$ hrs and $295$ km s$^{-1}$, $9.7$ hrs and $306$ km s$^{-1}$, and $11.4$ hrs and $307$ km s$^{-1}$, respectively.

Since the analyzed events cover an interval extending from 2008 until 2012, this study includes both periods of low and high solar activity. Forecasts are, however, more accurate during times of low solar activity. Table \ref{tab:split} shows the mean values and standard deviations of $\Delta t$ and $\Delta v$ of the ElEvoHI, FPF, HMF and SSEF forecasts for events 1--10 (2008--early 2011) and events 11--21 (2011--2012). Compared to solar maximum, we find the ElEvoHI arrival time forecast to be more accurate during low solar activity by about $\approx 2$ hrs. The arrival speed forecast shows the same behavior, were the difference between solar minimum and maximum is $\approx 50$ km s$^{-1}$. This behavior is even larger for FPF, HMF and SSEF---especially for the arrival speed forecasts.

 \cite{vrs14} compared the performance of the DBM model with that of the ``WSA-Enlil+Cone'' model \citep[][]{arp00,arg04,ods03}. They found that the latter yielded a mean arrival time error of $-0.3 \pm 16.9$ hrs. In order to make their DBM forecast, three different combinations of $\gamma$ and $w$ were used. DBM yielded a mean arrival time error of $1.9 \pm 18.8$ hrs, over all combinations. Splitting their sample of 50 CMEs  by solar activity, revealed (as we also see here) a smaller arrival time error during solar minimum conditions and a larger error during solar maximum.

For four CMEs in our event list (events 1, 3, 4, and 6), we find that the resulting drag parameter, $\gamma$, is higher than usual ($\approx 3\times10^{-7}$ km$^{-1}$). For one event, $\gamma=6.8\times10^{-7}$ km$^{-1}$, which may be a consequence of the DBM fit starting too close to the Sun. However, for this event, the fit does not converge if one assumes a later value for $t_{\rm init}$. Nevertheless, the forecasted arrival time and speed errors are not significantly worse than for other CMEs. There are several possible reasons why this fit might yield an \enquote*{unphysically} high value for the drag parameter. One such possibility may be the difference between the assumed and true background solar wind speed acting on the CME during its propagation. Note that the value of the solar wind speed ultimately used by ElEvo to provide time/speed estimates is that value (from the range delimited by the minimum and maximum solar wind speed over the course of the HI observations) that gives the best DBM fit.

\subsection{Relevance for prediction of geomagnetic effects}

Forecasting the arrival speed of CMEs plays a major role in the prediction of geomagnetic storm intensities. The strength of a geomagnetic storm is quantified through the use of several geomagnetic indices, one being the disturbance storm time ($Dst$) index. \citet{bur75} and \citet{obrmcp00} have developed models to derive $Dst$ from solar wind parameters, in particular the $B_z$ component of the magnetic field vector and the solar wind speed. Determining the magnetic field orientation within CMEs, prior to their arrival at Earth, particularly $B_z$, is one of the most important topics in space weather research and operational space weather. While such forecasting of $B_z$ is as yet unachievable in practice, ElEvoHI does appear to provide a reliable speed forecast that could be used to model $Dst$. In order to assess this approach, we have calculated the $Dst$ index for a CME (event number 20 in our list) that had a shock arrival speed of $\approx 500$ km s$^{-1}$, and that resulted in a moderate geomagnetic storm with a minimum $Dst$ of $-86$ nT.
We used the IMF $B_z$ component measured in situ by \textit{Wind} and a variety of different values of CME arrival speeds (within the errors of ElEvoHI and the FPF method) to model $Dst$ using the method of O'Brien \& McPherron. Use of the measured in situ arrival speed, results in a modeled minimum $Dst$ of $-96$ nT. Adding the mean arrival speed error over all CMEs, $\langle \Delta v \rangle=32$ km s$^{-1}$, from ElEvoHI with $f=1.2$, to the in situ CME speed for this interval, yields a minimum value for $Dst$ of $102$ nT; adding instead the standard deviation of $89$ km s$^{-1}$ gives $Dst=-116$ nT. Adding $\langle \Delta v \rangle=312$ km s$^{-1}$, the mean arrival speed error from FPF, to the in situ CME speed, gives $Dst=-145$ nT; adding the FPF standard deviation of 301 km s$^{-1}$ results in $Dst=-187$ nT. Correctly predicting CME arrival speed at Earth is of great importance for modeling the intensities of geomagnetic storms---this issue appears to be very well addressed using ElEvoHI. 

A very important factor for space weather forecasting is the prediction lead time. This is the time between when the prediction is performed until the impact of the CME. The prediction lead times of ElEvoHI for the CMEs under study lie in the range of $-26.4 \pm 15.3$ hrs, which is similar to that quoted by \citet[][]{moe14} as we use a subset of those events. Using a shorter HI track to extend the prediction lead time, would likely lead to some increase in forecasting errors. However, this still needs to be investigated.

\begin{deluxetable}{lrrrrr}
\tablecolumns{6}
\tablewidth{0pc}
\tabletypesize{\small}
\tablecaption{Mean of arrival time and arrival speed difference (calculated$-$observed).  \label{tab:split}}
\tablehead{
\colhead{} & \multicolumn{2}{c}{Events 1--10} &\colhead{}&\multicolumn{2}{c}{Events 11--20}\\
\cline{2-3} \cline{5-6}\\
\colhead{$f$} & \colhead{$\langle \Delta t \rangle [$hrs$]$}&\colhead{$\langle \Delta v \rangle [$km s$^{-1}]$}&\colhead{}&\colhead{$\langle \Delta t \rangle [$hrs$]$}& \colhead{$\langle \Delta v \rangle [$km s$^{-1}]$}}
\startdata
$0.8$ & $3.6 \pm 4.6 $ & $-0.8 \pm 53.8$& & $9 \pm 5.2 $ & $35.5 \pm 52.8$\\
$1$ & $4.4 \pm 4.7 $ & $-7.2 \pm 48.4$& &$6.9 \pm 6.9 $ & $49.6 \pm 68.8$\\
$1.2$ & $5.7 \pm 5.8 $ & $-15.1 \pm 49.7$& & $4.3 \pm 5.6 $ & $90.6 \pm 89.8$\\ 
\tableline
FPF & $-5.9 \pm 6.8 $ & $114 \pm 107$& & $-9.8 \pm 6.6 $ & $523 \pm 275$\\
HMF & $2.5 \pm 6.7 $ & $47 \pm 77$& & $-3.8 \pm 9.7 $ & $292 \pm 306$\\
SSEF & $5.3 \pm 8.8 $ & $30 \pm 82$& & $-2.3 \pm 11.4 $ & $277 \pm 307$\\
\enddata
\end{deluxetable}

\section{Discussion}

We have introduced the new HI-based CME forecasting utility, ElEvoHI, which assumes an elliptically shaped CME front, that adjusts, through drag, to the background solar wind speed during its propagation. Included within ElEvoHI is the newly presented conversion method, ElCon, which converts the time-elongation profile of the CME obtained from heliospheric imagery into a time-distance profile assuming an elliptical CME geometry; the resultant time-distance profile is used as input into the subsequent stage of ElEvoHI, the DBM fit. As a last stage of ElEvoHI, all of the resulting parameters are input in the Ellipse Evolution model \citep[ElEvo;][]{moe15}, which then predicts arrival times and speeds. 

We have assessed the efficacy of the new ElEvoHI procedure by forecasting the arrival times and speeds of 21 CMEs, previously analyzed by \citet{moe14}, which were detected in situ at 1~AU. In our implementation of ElEvoHI, the CME propagation directions were provided by the Fixed-$\phi$ fitting (FPF) method. ElEvoHI predictions of arrival times and speeds were compared to the output of other single-spacecraft HI-based methods, specifically FPF, Harmonic Mean fitting (HMF) and Self-similar Expansion fitting (SSEF).

We have found that ElEvoHI performs somewhat better at forecasting CME arrival time than the FPF, HMF and SSEF methods. Applying ElEvoHI with $f=0.8$, results in a mean error between predicted and observed arrival times of $\Delta t=6.4 \pm 5.3$ hrs; the equivalent values for FPF, HMF and SSEF are $\Delta t=-7.9 \pm 7$ hrs, $\Delta t=-3.8 \pm 9.7$ hrs and $\Delta t=-2.3 \pm 11.4$ hrs, respectively. Hence, while FPF, HMF and SSEF tend to forecast the CME arrival too early \citep[cf.][]{lugkin12}, ElEvoHI has a tendency to predict their arrival too late.
A substantial improvement is shown in forecasting the arrival speed. The mean error between the modeled and observed arrival speeds for ElEvoHI ($f=0.8$) is $16 \pm 53$ km s$^{-1}$, whereas for FPF, HMF and SSEF $\Delta v = 328 \pm 295$ km s$^{-1}$, $\Delta v = 292 \pm 306$ km s$^{-1}$ and $\Delta v = 277 \pm 306$ km s$^{-1}$, respectively. This improvement has a direct impact on the accuracy of predicting the intensity of geomagnetic storms at Earth.

\citet{sac15} demonstrated that the drag force is dominant at heliocentric distances of 15--50 R$_\odot$. Below this distance, the Lorentz force still influences CME kinematics. Our study supports this conclusion; we find DBM applicable beyond a mean heliocentric distance, $r_{\rm init}$, of $\approx 21 \pm 10$ R$_\odot$. As pointed out by \citet{har12}, it is quite likely that the FPF method (as well as HMF and SSEF) performs better if it is applied to data starting at a larger heliocentric distance as well.

ElEvoHI requires the presence of a heliospheric imager instrument providing a side view of the Sun-Earth line. These data must be available in near real-time, with an acceptable quality, to be able to use them for predicting CME arrival \citep[][]{tuc15}. The necessity of hosting heliospheric imagers at L4 and/or L5 is obvious when one compares the efficacy of using HI data to forecast CME arrival to methods using coronagraph data. For example, $66\%$ of the coronagraph-driven DBM and WSA-Cone model Enlil arrival time forecasts lie within the range of $\pm 18$ hrs \citep[][]{vrs14}. Using ElEvoHI, with the benefit of HI observations, this value is improved to around $\pm 6.5$ hrs.

Our study shows that there is no significant difference between the three aspect ratios used. A follow up study may discover the most appropriate curvature (and indeed angular width) to select by comparing the results to multiple longitudinally separated spacecraft detecting the CME arrival.

\acknowledgments

\acknowledgments{Support by the Austrian Science Fund (FWF): P26174-N27 is acknowledged by TR and CM. The presented work has received funding from the European Union Seventh Framework Programme (FP7/2007-2013) under grant agreement No.~606692 [HELCATS]. This research has made use of SunPy, an open-source and free community-developed solar data analysis package written in Python. We thank the \textit{STEREO SECCHI/IMPACT/PLASTIC} teams for their open data policy.}






\appendix

\section{Appendix}
\label{sec:appA}

\begin{figure}[!htbp]
\epsscale{.8}
\plotone{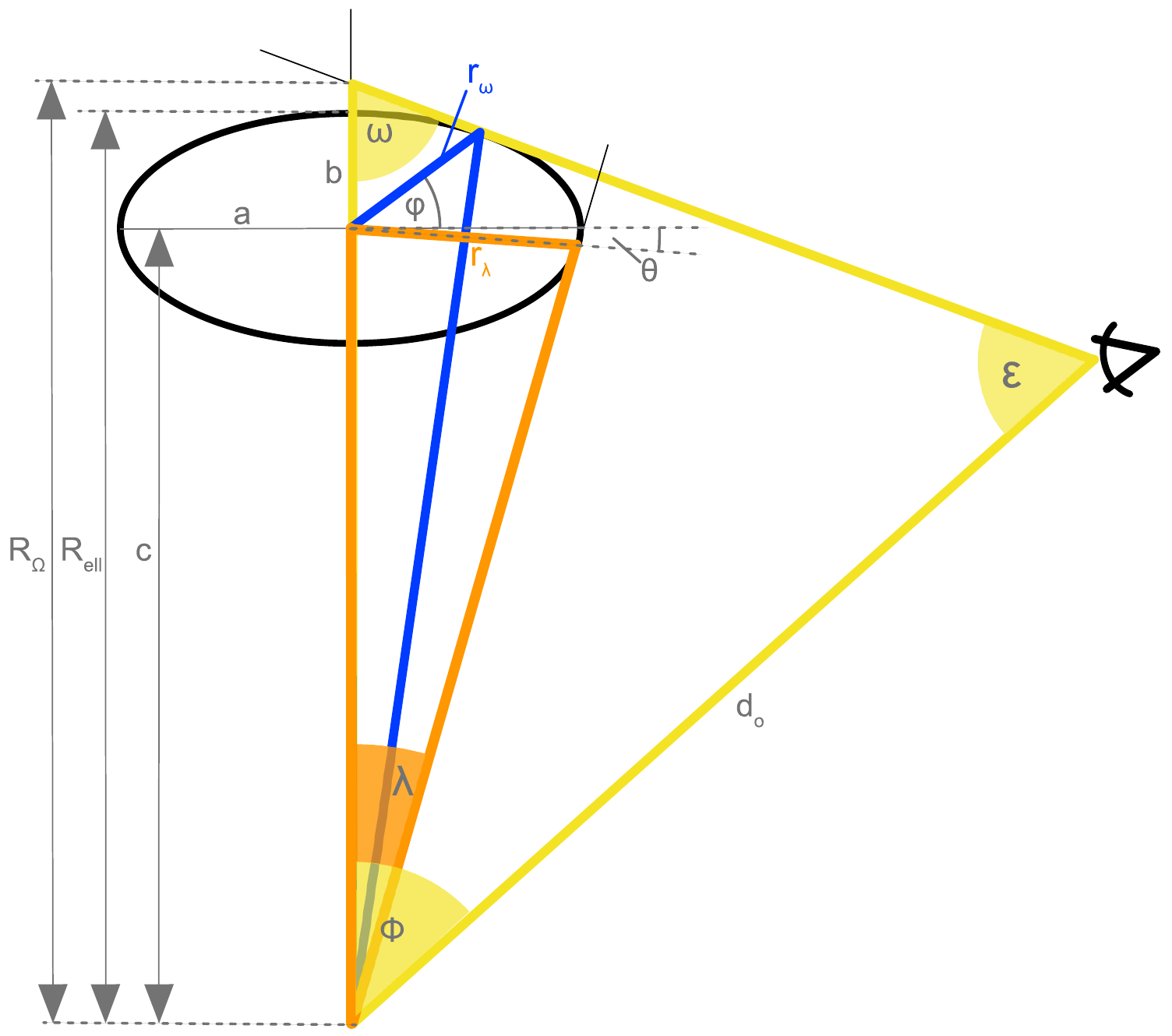}
\caption{Sketch of the elliptic geometry. We refer to the colored triangles to make the derivation more understandable. \label{fig:meth}}
\end{figure}

We use the ellipse conversion method (ElCon), to calculate the distance of the CME apex to Sun center, $R_{\rm ell}$, as a function of elongation, $\epsilon$, which is the angle between the Sun-observer line and the line of sight, assumed to be the tangent to the elliptical CME front. The elongation of the CME front, $\epsilon$, is available from STEREO/HI imagery. We also know the distance of the observer to the Sun, $d_{\rm o}$. The CME propagation direction, $\phi$, can be obtained by Fixed-$\phi$ fitting (see Section \ref{sec:input}), the inverse ratio of the semi-axes, $f=b/a$, and the CME half-width, $\lambda$, need to be assumed.
Figure \ref{fig:meth} illustrates the ElCon geometry.

\noindent By applying the sine rule to the yellow triangle we find

\begin{equation}
R_\Omega = \frac{d_{\rm o} \sin \epsilon}{\sin \omega}.
\label{eq:rOmega}
\end{equation}
Similarly, we find $R_\Omega - c$ is given as
\begin{equation}
R_\Omega - c = \frac{r_\omega \sin(90+\varphi-\omega)}{\sin \omega}.
\label{eq:rOmegamC}
\end{equation}

\noindent Using Equations \ref{eq:rOmega} and \ref{eq:rOmegamC}, and by applying the sine rule to the orange triangle, we can make the following ansatz:

\begin{equation}
  \begin{aligned}
     c &=\frac{d_{\rm o} \sin \epsilon - r_\omega \sin(90+\varphi-\omega)}{\sin \omega} \\
     c &=\frac{r_\lambda \sin(90+\theta - \lambda)}{\sin \lambda}
   \end{aligned}
\label{eq:ansatzc}
\end{equation}

\noindent The angles $\varphi$ and $\theta$ can be expressed as 
\begin{equation}
\varphi = \arctan(f^2 \tan \omega)
\label{eq:phi}
\end{equation}
and
\begin{equation}
\theta = \arctan(f^2 \tan \lambda).
\label{eq:omega}
\end{equation}

\noindent Furthermore, the distances of the two tangent points to the ellipse center, $r_\omega$ and $r_\lambda$, can be expressed as
\begin{equation}
r_\omega = \frac{b}{\sqrt{f^2 \cos^2\varphi+\sin^2 \varphi}}
\label{eq:rphi}
\end{equation}
and
\begin{equation}
r_\lambda = \frac{b}{\sqrt{f^2 \cos^2\theta+ \sin^2 \theta}}.
\label{eq:rlambda}
\end{equation}

\noindent By equating the expressions for $c$ in Equations \ref{eq:ansatzc} we can solve for the semi-minor axis as
\begin{equation}
b=\frac{d_{\rm o} \sin \epsilon \sin \lambda\, \Omega_{\theta} \, \Omega_{\varphi}}{\sin(90+\theta-\lambda) \sin \omega \,  \Omega_{\varphi}+\sin(90+\varphi-\omega) \sin \lambda \, \Omega_{\theta}}
\label{eq:b}
\end{equation}
with

\begin{equation}
     \Omega_x =\sqrt{f^2 \cos^2 x + \sin^2 x},  \quad x \in \{\theta, \varphi \}.\\
\end{equation}

\noindent Note that $\omega=180^\circ - \epsilon-\phi$. In the case of $\epsilon+\phi < 90^\circ$, we have to replace $\omega$ by $\omega=\epsilon+\phi$.

\vspace{0.5cm}

\noindent Now we are able to calculate $c$ using the second Equation in ansatz \ref{eq:ansatzc} and find the distance $R_{\rm ell}$ of the CME apex from Sun center:
\begin{equation}
R_{\rm ell}=c+b.
\label{eq:rell}
\end{equation}

To calculate the arrival time of the CME front at any location in interplanetary space, we need to account for the offset between the direction of the CME apex and the direction of the location of interest, the so-called off-axis correction. The distance from Sun center of the CME front at an angular offset $\Delta$ from its axis, $R_{\rm is}$, as presented by \citet{moe15} in their Equation 12, is given by

\begin{equation}
 R_{\rm is}=\frac{c \cos\Delta + \sqrt{(b^2-c^2) f^2 \sin^2\Delta+b^2 \cos^2\Delta}}{f^2 \sin^2\Delta+\cos^2\Delta}.
 \label{eq:ris}
 \end{equation}


\let\cleardoublepage\clearpage

\section{Appendix B}
\label{sec:appB}

\begin{deluxetable}{lclrrrc}
\tablecolumns{7}
\tablewidth{0pc}
\tabletypesize{\scriptsize}
\tablecaption{List of input parameters obtained from ElCon and DBM fitting, which are used for ElEvo. \label{tab:taball}}
\tablehead{
\colhead{Event}  & \colhead{} & \multicolumn{5}{c}{ElCon and DBM fit} \\
\cline{1-1} \cline{3-7} \\
\colhead{n\textordmasculine} & \colhead{} & \colhead{$t_{\rm init}$} & \colhead{$r_{\rm init}$} & \colhead{$v_{\rm init}$} & \colhead{w} & $\gamma$\\
\colhead{}  &\colhead{}  & \colhead{ {\tiny[UT]}} & \colhead{ {\tiny$R_\odot$}} & \colhead{{\tiny[km s$^{-1}$]}} & \colhead{{\tiny[km s$^{-1}$]}} & \colhead{{\tiny[$10^{-7}$km$^{-1}$]}}}
\startdata
$f=0.8$& & & & & & \\
\tableline
1 & & 2008 Apr 26 23:05 & 30.2 & 953 & 556 & 3.06 \\
2 &  & 2008 Jun 2 14:38 & 21.2 & 373 & 424 & 2.5\\
3& & 2008 Dec 27 10:34 & 10.6 & 714 & 404 & 2.6\\
4&  & 2009 Feb 13 08:32 & 6.9 & 373 & 292 & 3.07 \\
5 &  & 2010 Apr 3 16:42 & 35.8 & 1145 & 589 &0.76 \\
6& & 2010 Apr 8 04:01 & 3.6 & 989 & 480 &6.83 \\
 7 & & 2010 May 24 02:08 & 22 & 465 & 375 & 0.93 \\
 8 &  & 2010 Jun 16 22:39 & 18.6 & 358 & 492 & 0.22 \\
9 & & 2010 Aug 1 15:34 & 40.3 & 810 & 526 & 0.86 \\
10 & & 2011 Feb 15 03:11 & 11.7 & 847  & 516 & 1.43 \\
11 & & 2011 Aug 2 10:34 & 21.8 & 562 & 377 & 0.23 \\
12 & & 2011 Sep 7 05:12 & 25.8 & 608 &  361 & 0.04 \\
13 & & 2011 Oct 22 09:40 & 27.1 & 669 & 352 &0.21 \\
 14 & & 2012 Jan 19 19:28 & 19.8 & 889 & 294 &0.23 \\
15 & & 2012 Jan 23 07:33 & 39.6 & 2237 & 470 & 0.43 \\
16& & 2012 Mar 5 07:39 & 17.4 & 1086 & 406 & 0.41 \\
17 & & 2012 Mar 7 02:00 & 16.6 & 1480  & 606 &0.26 \\
18 & & 2012 Mar 10 19:17 & 12.3 & 1883 & 489 & 0.45 \\
19 & & 2012 Apr 19 19:05 & 13.3 & 815 &  366 & 0.75 \\
20 & & 2012 Jun 14 15:49 & 15.4 & 1349 &  381 &0.37 \\
21 & & 2012 Jul 12 20:47 & 26.2 & 1079 &  373 &0.14 \\
\tableline
$f=1$& & & & & & \\
\tableline
1 & & 2008 Apr 26 23:05 & 30.3 & 952 & 556 & 3.08 \\
2 & & 2008 Jun 2 14:38 & 21.7 & 379 & 424 &1.63\\
3 & & 2008 Dec 27 10:34 & 10.6 & 717 &  404 & 2.01 \\
4& & 2009 Feb 13 14:34 & 19.8 & 355 &  292 &3.31 \\
5& & 2010 Apr 3 16:06 & 35.8 & 1143 &  589 &0.7 \\
6 & & 2010 Apr 8 04:01 & 3.6 & 989 & 480 &5.99 \\
7 & & 2010 May 24 02:08 & 22.2 & 468  & 375 &1.01 \\
8& & 2010 Jun 16 22:39 & 18.8 & 360 &  492 &0.18 \\
9 & & 2010 Aug 1 15:34 & 40.5 & 822 & 526 &0.55 \\
10 & & 2011 Feb 15 05:12 & 20.7 & 720 & 516 &0.72 \\
11 & & 2011 Aug 2 10:34 & 22.0 & 570 &  377 &0.12 \\
12& & 2011 Sep 7 10:44 & 38.1 & 1448 &  423 &1.59 \\
13& & 2011 Oct 22 09:40 & 27.2 & 674 & 352 &0.15 \\
14 & & 2012 Jan 19 19:28 & 20.2 & 908 & 294 &0.18 \\
15 & & 2012 Jan 23 07:33 & 40.4 & 2318 & 470 &0.37\\
16& & 2012 Mar 5 07:39 & 17.8 & 1114 & 406 &0.32\\
17 & & 2012 Mar 7 02:00 & 17.8 & 1519 & 606 &0.19 \\
18 & & 2012 Mar 10 19:17 & 12.3 & 1888 & 489 &0.41 \\
19 & & 2012 Apr 19 20:36 & 18.8 & 625 & 348 &0.16 \\
20 & & 2012 Jun 14 16:20 & 15.6 & 1438 &  381 &0.46 \\
21 & & 2012 Jul 12 17:45 & 6.9 & 1396 & 422 &0.23 \\
\tableline
$f=1.2$& & & & & & \\
\tableline
1 & & 2008 Apr 26 23:05 & 30.3 & 951 & 556 &3.10 \\
2& & 2008 Jun 2 14:38 & 22.1 & 383 & 424 & 1.03 \\
3& & 2008 Dec 27 10:34 & 10.6 & 719 & 454 &4.89 \\
4 & & 2009 Feb 13 14:34 & 19.8 & 357 & 292 &2.09 \\
5 & & 2010 Apr 3 15:06 & 26.7 & 2052 & 589 &1.26 \\
6 & & 2010 Apr 8 04:01 & 3.6 & 989 & 480 &5.51 \\
7& & 2010 May 24 02:08 & 22.4 & 469 & 375 &1.07 \\
8& & 2010 Jun 16 22:39 & 18.9 & 361 & 492 &0.17 \\
9 & & 2010 Aug 1 15:34 & 40.7 & 830 & 526 &0.39 \\
10 & & 2011 Feb 15 05:12 & 20.8 & 725 & 516 &0.49 \\
11 & & 2011 Aug 2 10:34 & 22.1 & 574 & 377 &0.06\\
12 & & 2011 Sep 7 05:12 & 26.1 & 623 & 391 &0.06\\
13 & & 2011 Oct 22 11:41 & 34.4 & 738 & 352 &0.21 \\
14& & 2012 Jan 19 19:28 & 20.4 & 921 & 294 &0.15\\
15 & & 2012 Jan 23 07:33 & 40.9 & 2377 & 470 &0.33\\
16& & 2012 Mar 5 05:39 & 9.1 & 1054 & 406 &0.19\\
17& & 2012 Mar 7 02:00 & 17.2 & 1546 & 606 &0.15 \\
18& & 2012 Mar 10 19:17 & 12.3 & 1892 & 489 &0.38 \\
19 & & 2012 Apr 19 19:05 & 13.5 & 832 & 366 &0.53 \\
20 & & 2012 Jun 14 15:49 & 15.8 & 1399 &  381 &0.27 \\
21& & 2012 Jul 12 17:45 & 6.9 & 1409 & 422 &0.2\\
\enddata
\tablenotetext{*}{FPF results taken from \citet{moe14}.}
\end{deluxetable}

\cleardoublepage




\end{document}